\documentclass[compsoc,conference,a4paper,10pt,times]{IEEEtran}
\IEEEoverridecommandlockouts
\usepackage{cite}
\usepackage{amsmath,amssymb,amsfonts}
\usepackage{algorithmic}
\usepackage{graphicx}
\usepackage{textcomp}
\usepackage{bmpsize}
\usepackage{xcolor}
\usepackage[utf8]{inputenc}
\usepackage{lipsum}
\usepackage[colorlinks=true,urlcolor=black]{hyperref}

\usepackage{todonotes}
\usepackage{tikz}
\usetikzlibrary{positioning}

\pgfdeclarelayer{bg}    
\pgfsetlayers{bg,main}  

\def\BibTeX{{\rm B\kern-.05em{\sc i\kern-.025em b}\kern-.08em
    T\kern-.1667em\lower.7ex\hbox{E}\kern-.125emX}}
\begin{document}

\title{Password similarity using probabilistic data structures}


\author{
	\IEEEauthorblockN{Davide Berardi, Franco Callegati, Andrea Melis, Marco Prandini}
	\IEEEauthorblockA{\textit{Dipartimento di Informatica Scienza e Ingegneria (DISI)} \\
	\textit{Università di Bologna}\\
	Bologna, Italy\\
	\{{davide.berardi6, franco.callegati, a.melis, marco.prandini}\}@unibo.it}
}

\maketitle

\begin{abstract}
Passwords should be easy to remember, yet expiration policies mandate their frequent change.
Caught in the crossfire between these conflicting requirements, users often adopt creative
methods to perform slight variations over time. While easily fooling the most basic checks for
similarity, these schemes lead to a substantial decrease in actual security, because leaked
passwords, albeit expired, can be effectively exploited as seeds for crackers.
This work describes an approach based on Bloom filters to detect password similarity, which 
can be used to discourage password reuse habits. The proposed scheme intrinsically obfuscates 
the stored passwords to protect them in case of database leaks, and can be tuned to be resistant
to common cryptanalytic techniques, making it suitable for usage on exposed systems.
\end{abstract}


\section{Introduction}
Text-based passwords are still the most common way to authenticate users against
services\cite{schneier2005two}. According to a typical classification for authorization mechanisms,
they fall in the ``what you know'' category. The other categories, ``what
you are'' and ``what you have'', are most commonly used only as a second factor 
for the so-called \emph{MFA} multiple factor
authentication schemes\cite{scheidt2007access}. An example of MFA is a 
one time password available only on a personal device (e.g.
smartphone) at each login attempt into a critical service. 
Passwords should be easy to remember, but hard to guess.
In a long game of cops and robbers, users try to base them 
on dictionary words\cite{stobert2014password}, and system administrators
write policies to block these attempts. As a common reaction, users make
simple variations to hide the base, easy-to-remember common word. This 
makes the password harder to remember, so users try to stick to the same one
forever, but system administrators add expiration times to their policies. 
Users then adapt by making the smallest possible change at each password update.
This is still an insecure behavior, making life much easier to malicious 
actors trying to guess current passwords, because old ones are often available
through leaked databases.
The countermeasure to discourage this behaviour is to prevent choosing 
a password too similar to the previous one(s). The problem with the simplest approaches
to similarity estimation — for example based on the Levenshtein
distance\cite{levenshtein1966binary} between two
clear-text strings — is that passwords are sensitive and personal data.
Storing them to enable similarity checks exposes users to unacceptable risks 
of breaches that may result in unauthorized access to some of their accounts.
To overcome this problem, in this paper we propose a system to calculate the 
similarity of passwords over obfuscated data, based on Bloom filters. 
The tuning of the filter is thoroughly analyzed to determine the values of parameters
that ensure password secrecy, yet allow achieving an effective detection of similarity. 
The scheme natively allows the integration of a cryptographic access control method,
which will be investigated in a future work.
The paper is structured in the following way: section \ref{sec:state_of_the_art} 
discusses a list of previous works and state-of-the-art tools that employ
similar techniques to evaluate similarity between data. The details of
password similarity, design of our system and a possible attack on this kind of data
analyzers are introduced in section \ref{sec:password}. 
Our scheme was implemented in C language for the Linux operating system, and a performance analysis 
is presented in section \ref{sec:experimental}.
We illustrate some application scenarios and propose
future improvements for our work in section \ref{sec:application}.
\section{State of the art}
\label{sec:state_of_the_art}
The foundations for this work can be traced back to Schnell's (et alii) paper
\cite{schnell2009privacy}, which describes a method for querying
private data over a Bloom filter structure. Another work presented in
\cite{alaggan2012blip} describes the application of privacy methods like
differential privacy to probabilistic data structures such as Bloom
filters. While the described approach is vulnerable to an attack called the ``Profile
Reconstruction Attack'', this is due to use of differential privacy methods 
(which are not exploited in our work) and not to the filter itself. Anonymized datasets, using techniques such as differential privacy,
purposely introduces errors and noise in the data in order to hide the presence of specific information or to ensure that links between users and his correspondant data cannot be established.
This controlled errors are compensated in large dataset: the errors do not effect the quality of such evaluations. In our use cases, users - even if forced to change password regularly - can't generate a password dataset big enough to compensate the introduced noise. That is, anonimization techniques can affect
the quality of password similarity queries, with several false positives compared to the proposed bloom filter approach. We still did not provide an experimental analysis of this statement which is planned to be implemented in future works.
%
RAPPOR\cite{erlingsson2014rappor} is a system used by Google to get data
from the Chrome browser. The data is hashed in a Bloom filter,
anonymized by introducing a perturbation on the values, and then retrieved and
reconstructed at server side. This approach is similar to the one we
propose in this work, but it explicitly does not apply to passwords, to avoid
sending potentially sensitive data from the device to a cloud system.
To the best of our knowledge, in the literature there is not a use case 
for a password scenario. For this reason the aim of this paper is to study it, illustrate
the advantages it brings, and discuss the security-related issues that it introduces.
\section{Password similarity}
\label{sec:password}
It is widely known that password reuse
is a common behaviour which can turn into a threat if passwords get leaked
online\cite{ives2004domino}. Password leaks are a common form of
information leak that happens regularly\footnote{https://us.norton.com/internetsecurity-emerging-threats-2019-data-breaches.html}. 
%
A similar threat appears when
a user is forced to a password change and (especially for corporate passwords that
must be changed quite often), he can insert a different password
with little variations from the last one (e.g. {\texttt~password2020} changing
password from {\texttt~password2019}).  This can make brute force attacks 
very effective, since the new password is easily computed by a limited number
of mutations starting from a dictionary of leaked ones.
%
We can describe the password mutation as a perturbation of the password
with slight variants.  Tools such as password crackers or word-lists
generators like cupp\footnote{A password generator based on personal and
open source data: \url{https://github.com/Mebus/cupp}} or
johntheripper\footnote{Johtheripper is a common password cracker and
generator \url{https://www.openwall.com/john/}} implements various methods for
password generation, the passwords can also be generated by neural-based
techniques such as adversarial generation\cite{liu2018genpass}.  With
these approaches, choosing a similar password is almost as insecure as choosing
a dictionary password\cite{wood1996constructing}.
For the purpose of this work, password similarity can be informally defined as the structural similarity
of the text composing two password being compared, i.e. it has nothing to do
with the possible underlying meaning of the string. This definition of password similarity 
can be used to guarantee that a user is not
reciclying what, in terms of actual entropy, can be considered the same password every time.
Common methods to guide users to choose ``robust'' passwords are focused on avoiding 
the direct use of dictionary words. These methods have two shortcomings: users resource to variations
that are easily discovered by mutation by the aforementioned tools, and there is no
detection of password similarity when a password change is mandated.
A straightforward method to detect password similarity over a meaningful time span would 
require saving old passwords in clear-text into a database, which is obviously a despicable approach.
We propose a solution based on Bloom filters that overcomes this shortcoming.
\subsection{Bloom filters for text similarity}
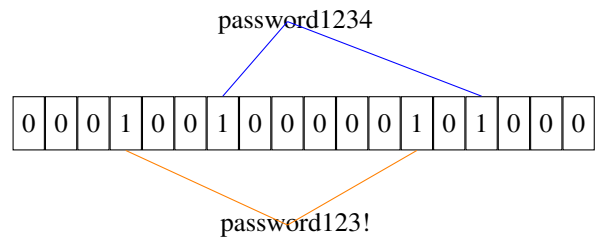
\begin{figure}[h]
	\centering
	\begin{tikzpicture}
		\node[draw,minimum width=1em, minimum height=2em] (bf1) {0};
		\node[draw,minimum width=1em, minimum height=2em,right=0cm of bf1] (bf2) {0};
		\node[draw,minimum width=1em, minimum height=2em,right=0cm of bf2] (bf3) {0};
		\node[draw,minimum width=1em, minimum height=2em,right=0cm of bf3] (bf4) {1};
		\node[draw,minimum width=1em, minimum height=2em,right=0cm of bf4] (bf5) {0};
		\node[draw,minimum width=1em, minimum height=2em,right=0cm of bf5] (bf6) {0};
		\node[draw,minimum width=1em, minimum height=2em,right=0cm of bf6] (bf7) {1};
		\node[draw,minimum width=1em, minimum height=2em,right=0cm of bf7] (bf8) {0};
		\node[draw,minimum width=1em, minimum height=2em,right=0cm of bf8] (bf9) {0};
		\node[draw,minimum width=1em, minimum height=2em,right=0cm of bf9] (bf10) {0};
		\node[draw,minimum width=1em, minimum height=2em,right=0cm of bf10] (bf11) {0};
		\node[draw,minimum width=1em, minimum height=2em,right=0cm of bf11] (bf12) {0};
		\node[draw,minimum width=1em, minimum height=2em,right=0cm of bf12] (bf13) {1};
		\node[draw,minimum width=1em, minimum height=2em,right=0cm of bf13] (bf14) {0};
		\node[draw,minimum width=1em, minimum height=2em,right=0cm of bf14] (bf15) {1};
		\node[draw,minimum width=1em, minimum height=2em,right=0cm of bf15] (bf16) {0};
		\node[draw,minimum width=1em, minimum height=2em,right=0cm of bf16] (bf17) {0};
		\node[draw,minimum width=1em, minimum height=2em,right=0cm of bf17] (bf18) {0};

		\node[above=2em of bf9] (pass1) {\texttt~password1234};

		\node[below=2em of bf9] (pass2) {\texttt~password123!};

		\draw[-,blue] (pass1.center) -- (bf7.north) {};
		\draw[-,blue] (pass1.center) -- (bf15.north) {};

		\draw[-,orange] (pass2.center) -- (bf4.south) {};
		\draw[-,orange] (pass2.center) -- (bf13.south) {};
	\end{tikzpicture}
	\caption{Insertion procedure with two strings.  The strings
	$password1234$ and $password123!$ are hashed
	independently.  This Insertion procedure process the passwords
	as single items, leading to different hashed values.}
	\label{fig:insert}
\end{figure}

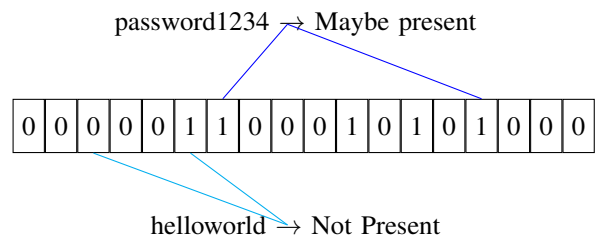
\begin{figure}[h]
	\centering
	\begin{tikzpicture}
		\node[draw,minimum width=1em, minimum height=2em] (bf1) {0};
		\node[draw,minimum width=1em, minimum height=2em,right=0cm of bf1] (bf2) {0};
		\node[draw,minimum width=1em, minimum height=2em,right=0cm of bf2] (bf3) {0};
		\node[draw,minimum width=1em, minimum height=2em,right=0cm of bf3] (bf4) {0};
		\node[draw,minimum width=1em, minimum height=2em,right=0cm of bf4] (bf5) {0};
		\node[draw,minimum width=1em, minimum height=2em,right=0cm of bf5] (bf6) {1};
		\node[draw,minimum width=1em, minimum height=2em,right=0cm of bf6] (bf7) {1};
		\node[draw,minimum width=1em, minimum height=2em,right=0cm of bf7] (bf8) {0};
		\node[draw,minimum width=1em, minimum height=2em,right=0cm of bf8] (bf9) {0};
		\node[draw,minimum width=1em, minimum height=2em,right=0cm of bf9] (bf10) {0};
		\node[draw,minimum width=1em, minimum height=2em,right=0cm of bf10] (bf11) {1};
		\node[draw,minimum width=1em, minimum height=2em,right=0cm of bf11] (bf12) {0};
		\node[draw,minimum width=1em, minimum height=2em,right=0cm of bf12] (bf13) {1};
		\node[draw,minimum width=1em, minimum height=2em,right=0cm of bf13] (bf14) {0};
		\node[draw,minimum width=1em, minimum height=2em,right=0cm of bf14] (bf15) {1};
		\node[draw,minimum width=1em, minimum height=2em,right=0cm of bf15] (bf16) {0};
		\node[draw,minimum width=1em, minimum height=2em,right=0cm of bf16] (bf17) {0};
		\node[draw,minimum width=1em, minimum height=2em,right=0cm of bf17] (bf18) {0};

		\node[above=2em of bf9] (pass1) {\texttt~password1234 $\to$ Maybe present};

		\node[below=2em of bf9] (pass2) {\texttt~helloworld $\to$ Not Present};

		\draw[-,blue] (pass1.center) -- (bf7.north) {};
		\draw[-,blue] (pass1.center) -- (bf15.north) {};

		\draw[-,cyan] (pass2.center) -- (bf6.south) {};
		\draw[-,cyan] (pass2.center) -- (bf3.south) {};
	\end{tikzpicture}
	\caption{Check procedure with two strings.  The strings $password1234$ and $helloworld$ are hashed
	independently and the resultant indexes from the hashe functions are checked in the bucket. If the
	lookup lead to a $0$ value, the string is not present in the filter.  Otherwise, it can be a value
	that is present in the filter or can be a collision (a false positive).}
	\label{fig:check}
\end{figure}

A Bloom filter is a probabilistic data structure which can be used to cache data
and speed up operations such as lookup in databases\cite{bloom1970space,mitzenmacher2002compressed}.
It is composed by:
\begin{itemize}
	\item a bucket which can be an array of bits initially set to the false value ($0$);
	\item a set of hash functions which will be used to insert and check values.
\end{itemize}
An insertion operation ($Insert(\beta,s)$) of a value in a Bloom filter is performed according to the following steps:
\begin{itemize}
	\item the value that must be inserted into the bucket is hashed using the set of hash functions;  The hash functions output must be re-mapped to provide indexes in the co-domain of cardinality $\kappa$.
	\item every bucket slot indexed by the keys got using the hash functions is set to the true value ($1$).
\end{itemize}
This operation therefore insert the hashed value into the filter, setting the correspondent hash values to the true value.  This is described in figure \ref{fig:insert} which pictures an insertion of the strings $password1234$ and $password123!$.
The verification process of the value presence in the filter ($Check(\beta,s) \to Boolean$) is analogous to the insertion case:
\begin{itemize}
	\item The element which is checked is hashed against all the functions to get a list of indexes;
    \item If any index points to a false value, then the element is not present in the filter for sure. The Bloom filter never exhibits false negatives.
	\item Otherwise the value can be present in the filter, but due to the collision possibility of the hash functions, the result can be a false positive.
\end{itemize}
This procedure is described in figure \ref{fig:check}.
Bloom filters can be used for text similarity using an n-gram approach. This techniques divides the string
in n-grams and hashes every resultant n-gram with the hash functions present in the set:
\begin{figure}[h]
	\centering
	\begin{tikzpicture}
		\node[draw,minimum width=1em, minimum height=2em] (bf1) {0};
		\node[draw,minimum width=1em, minimum height=2em,right=0cm of bf1] (bf2) {1};
		\node[draw,minimum width=1em, minimum height=2em,right=0cm of bf2] (bf3) {1};
		\node[draw,minimum width=1em, minimum height=2em,right=0cm of bf3] (bf4) {1};
		\node[draw,minimum width=1em, minimum height=2em,right=0cm of bf4] (bf5) {0};
		\node[draw,minimum width=1em, minimum height=2em,right=0cm of bf5] (bf6) {1};
		\node[draw,minimum width=1em, minimum height=2em,right=0cm of bf6] (bf7) {1};
		\node[draw,minimum width=1em, minimum height=2em,right=0cm of bf7] (bf8) {0};
		\node[draw,minimum width=1em, minimum height=2em,right=0cm of bf8] (bf9) {0};
		\node[draw,minimum width=1em, minimum height=2em,right=0cm of bf9] (bf10) {1};
		\node[draw,minimum width=1em, minimum height=2em,right=0cm of bf10] (bf11) {1};
		\node[draw,minimum width=1em, minimum height=2em,right=0cm of bf11] (bf12) {1};
		\node[draw,minimum width=1em, minimum height=2em,right=0cm of bf12] (bf13) {1};
		\node[draw,minimum width=1em, minimum height=2em,right=0cm of bf13] (bf14) {0};
		\node[draw,minimum width=1em, minimum height=2em,right=0cm of bf14] (bf15) {1};
		\node[draw,minimum width=1em, minimum height=2em,right=0cm of bf15] (bf16) {0};
		\node[draw,minimum width=1em, minimum height=2em,right=0cm of bf16] (bf17) {0};
		\node[draw,minimum width=1em, minimum height=2em,right=0cm of bf17] (bf18) {1};

		\node[above=2em of bf5] (pass1ngram1) {$pa$};
		\node[above=2em of bf7] (pass1ngram2) {$ss$};
		\node[above=2em of bf9] (pass1ngram3) {$wo$};
		\node[above=2em of bf11] (pass1ngram4) {$rd$};
		\node[above=2em of bf13] (pass1ngram5) {$12$};
		\node[above=2em of bf15] (pass1ngram6) {$34$};
		\node[above=4em of bf9] (pass1) {\texttt~password1234};

		\node[below=2em of bf5] (pass2ngram1) {$pa$};
		\node[below=2em of bf7] (pass2ngram2) {$ss$};
		\node[below=2em of bf9] (pass2ngram3) {$wo$};
		\node[below=2em of bf11] (pass2ngram4) {$rd$};
		\node[below=2em of bf13] (pass2ngram5) {$12$};
		\node[below=2em of bf15] (pass2ngram6) {$3!$};
		\node[below=4em of bf9] (pass2) {\texttt~password123!};

		\draw[-,blue] (pass1ngram1.center) -- (bf7.north) {};
		\draw[-,blue] (pass1ngram1.center) -- (bf15.north) {};

		\draw[-,orange] (pass1ngram2.center) -- (bf4.north) {};
		\draw[-,orange] (pass1ngram2.center) -- (bf13.north) {};

		\draw[-,cyan] (pass1ngram3.center) -- (bf6.north) {};
		\draw[-,cyan] (pass1ngram3.center) -- (bf3.north) {};

		\draw[-,pink] (pass1ngram4.center) -- (bf15.north) {};
		\draw[-,pink] (pass1ngram4.center) -- (bf10.north) {};

		\draw[-,gray] (pass1ngram5.center) -- (bf10.north) {};
		\draw[-,gray] (pass1ngram5.center) -- (bf12.north) {};

		\draw[-,red] (pass1ngram6.center) -- (bf18.north) {};
		\draw[-,red] (pass1ngram6.center) -- (bf11.north) {};

		\draw[-,blue] (pass2ngram1.center) -- (bf7.south) {};
		\draw[-,blue] (pass2ngram1.center) -- (bf15.south) {};

		\draw[-,orange] (pass2ngram2.center) -- (bf4.south) {};
		\draw[-,orange] (pass2ngram2.center) -- (bf13.south) {};

		\draw[-,cyan] (pass2ngram3.center) -- (bf6.south) {};
		\draw[-,cyan] (pass2ngram3.center) -- (bf3.south) {};

		\draw[-,pink] (pass2ngram4.center) -- (bf15.south) {};
		\draw[-,pink] (pass2ngram4.center) -- (bf10.south) {};

		\draw[-,gray] (pass2ngram5.center) -- (bf10.south) {};
		\draw[-,gray] (pass2ngram5.center) -- (bf12.south) {};

		\draw[-,green] (pass2ngram6.center) -- (bf12.south) {};
		\draw[-,green] (pass2ngram6.center) -- (bf2.south) {};
	\end{tikzpicture}
	\caption{n-gram insertion procedure with two strings.  The
	strings are divided into $n-grams$ (in this case bi-grams) and
	hashed using an $Insert$ operation for every $n-gram$.}
	\label{fig:qinsert}
\end{figure}
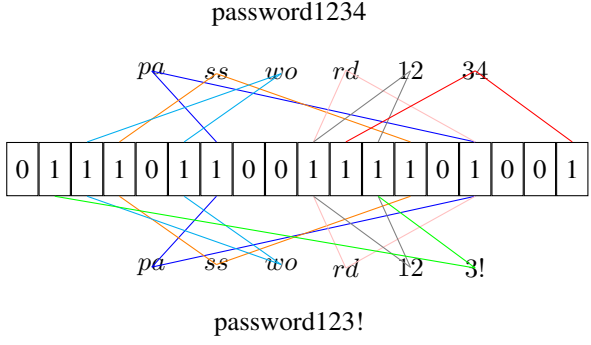
The hashing procedure to enable the measure of distance is presented in
figure \ref{fig:qinsert}.
By extending the check procedure to get the number of n-grams which are
the same in the two strings, independently of the order, the similarity of
the two sets can be calculated by using various distance
definitions. The similarity distance of two Bloom filters is commonly
expressed using the Jaccard coefficient\cite{brown2019evaluation} and is defined as:
\begin{equation}
	\delta(\beta_1,\beta_2) = \dfrac{2\gamma_{\beta_1,\beta_2}}{k_{\beta_1} + k_{\beta_2}}
\end{equation}
With $\gamma_{\beta_1,\beta_2}$ as the common number of true values in the sets of the two
Bloom filters $\beta_1$ and $\beta_2$.  $k_{\beta_1}$ and $k_{\beta_2}$ are the number of
true values of, respectively, the $\beta_1$ filter and $\beta_2$ filter.
Therefore the relevant operations on a filter to measure similarity are:
\begin{itemize}
	\item $Create(\Gamma,\kappa) \to \beta$ which generate a Bloom filter $\beta$
		using the hash functions present in the set $\Gamma$ with a
		bucket of size $\kappa$.
	\item $Insert(\beta,s)$ which insert the value $s$ in the Bloom filter.
	\item $Check(\beta,s) \to Boolean$ which check if the value $s$ is not present in the
		filter or if it collides with a present value.
	\item $QInsert(\beta,s,\nu)$ that insert the string $s$ splitting it in $\nu$-grams.
	\item $Distance(\beta_1,\beta_2) \to Real$ that returns the distance between two Bloom filters.
		To be comparable, two Bloom filters must have the same bucket size $\kappa$, and need to use the same set of hash functions $\Gamma$.
\end{itemize}
%

The effectiveness of the filter is strictly tied to the choice
of the hash function set $\Gamma$ and the size of the bucket $\kappa$.  
The tuning of these parameters is essential to achieve a satisfactory trade-off 
between the utility of the query and the number of false positives.
A wrong sizing of buckets or a choice of low-randomness hash functions can easily lead to a vulnerable filter (as detailed in section \ref{sec:anagramatk}) or to an unstable filter that leads to too many false positives.

\subsection{Privacy guarantees}
To size the filter there are some criteria which
can be derived from the following formulas:
\begin{equation}
	fpp = \left(1 -\left( 1 - \dfrac{1}{m}\right)^{kn}\right)^k
\end{equation}
Where $fpp$ is the false positive probability of
the filter.  This can be calculated a-priori from the variables $m$, $k$ and $n$, 
defined as:
\begin{itemize}
	\item $m$, the cardinality of the set on which the filter is built;
	\item $k$, the number of different hash functions that are used to
		hash values into the filter;
	\item $n$, the number of elements which will be inserted into the filter.
\end{itemize}
From this formula we can derive the optimal values for the filter size and the
number of hash functions to use for a specific number of elements that will be
inserted in the filter. The optimal value for $m$ can be calculated as:
\begin{equation}
m' = \left\lceil{- \dfrac{n\ln fpp}{(ln 2)^2}}\right\rceil
\label{eq:optimalm}
\end{equation}
The optimal number of
different hash functions is derived from the optimal size of the Bloom filter
internal set:
\begin{equation}
k' = \left\lceil{\dfrac{m'}{n}\ln2}\right\rceil
\end{equation}
with $m'$ being the optimal value for $m$ calculated as in \ref{eq:optimalm}.
The rationale behind this sizing formula comes from the observation that as
the size of buckets increases, the probability of collisions
decreases.  Using this approach, a Bloom filter with a controllable number
of false positives can be tuned to fit any specific scenario.
Conversely, as it is useful for the proposed application in order to 
enhance password confidentiality, a Bloom filter can
be designed in such a way to have a big number of false positives,
thus obfuscating data by forcing collisions.
Password confidentiality could be protected also by adopting privacy-preserving 
approaches, such as $\delta$-presence\cite{nergiz2007hiding}
or differential privacy\cite{dwork2008differential}, exist.  These approaches
take in consideration the amount of data stored into a database, or the
filter in this case, and try to anonymize the data between many false
positives.  In this case the approach is directly applicable to the filter
which can gain a lot of advantages in terms of privacy from this approach.
\subsection{Analysis of the hash function family}
Another key component of the filter which impacts on confidentiality is the
distribution of the values returned by the chosen hash function family.  In this
kind of probabilistic data structures, it may be necessary to deploy a huge
number of hash functions. Using a different algorithm for every index can be
unfeasible, and in any case a large variability within the hash function set 
hinders a precise analysis of the randomness of the generator.
To overcome these problems, the hash function set can be generated by using the
well-known salt approach, as described in figure \ref{fig:hashing}. 
When a value has to be inserted, many random numbers (salts) are generated. 
Prepending these numbers to the value, and using a fixed hash function, the 
effect is analogue to having adopted different, random hash functions.
In our implementation we used $MD5$ as the base hash function. The salts are
saved in the same location as the filter set, to allow reloading the state in
subsequent invocations of the algorithm. 
We acknowledge that this trivial solution is vulnerable to attacks
if the file is saved in clear-text form, like the one described in section
\ref{sec:anagramatk}. 

Saving salts in a more secure way  seems not too challenging, and is the subject of current and future investigation aimed at making the filter immune to this kind of attacks. For example, the salt can be generated using a cipher like AES using an user provided key 
and a fixed payload similarly to AES-CTR mode, extending the functions applicable on the
filter with a function $GenerateHashes(key) \to \Gamma$ which generate the set of hash
functions $\Gamma$ starting from the key $key$.
Using this technique, the filter set can 
be saved without specific protections, since it can be verified only using 
the chosen secret key. An in-depth analysis of the security of the encryption scheme,
particularly concerning the peculiarity of having very short payloads due to the division in n-grams, will be the focus of ongoing research work.
These approaches to the generation of hash functions are described in
figure \ref{fig:hashing}.  In the first hash set we can see that
$h1$ and $h2$ are two functions generated with a random padding applied
to the $MD5$ hash function.  In this scenario we suppose to have a
$Random(n)$ function that can generate a random string of length $n$.
This generation, when re-applied will lead to a totally different set of
hash functions, making the distance function inapplicable.
That is the concept of the construction of $h1'$ and $h2'$.  These functions can
return different results from the functions $h1$ and $h2$ described before.
Reusing the same value for the salt applied to the functions will lead to the
same set of results, making the distance calculation possible.  This is
the concept of the third figure, which describes how, applying a fixed
salt to a symmetric cipher using a secret key $k$ the filter will lead
to the same set of results.
\begin{figure}
	\begin{tikzpicture}
		\node[] (label0)                    {$h1(s)=MD5(Random(10) + s)$};
		\node[below=.5cm of label0.west, anchor=west] (label1) {$h2(s)=MD5(Random(10) + s)$};
		\node[below=.5cm of label1.west, anchor=west] (label2) {$Insert(\beta,'password1234') \to \{6,15\}$};

		\node[draw,minimum width=1em, minimum height=2em,below=.7cm of label2.west, anchor=west] (bf1) {0};
		\node[draw,minimum width=1em, minimum height=2em,right=0cm of bf1] (bf2) {0};
		\node[draw,minimum width=1em, minimum height=2em,right=0cm of bf2] (bf3) {0};
		\node[draw,minimum width=1em, minimum height=2em,right=0cm of bf3] (bf4) {0};
		\node[draw,minimum width=1em, minimum height=2em,right=0cm of bf4] (bf5) {0};
		\node[draw,minimum width=1em, minimum height=2em,right=0cm of bf5] (bf6) {0};
		\node[draw,minimum width=1em, minimum height=2em,right=0cm of bf6] (bf7) {1};
		\node[draw,minimum width=1em, minimum height=2em,right=0cm of bf7] (bf8) {0};
		\node[draw,minimum width=1em, minimum height=2em,right=0cm of bf8] (bf9) {0};
		\node[draw,minimum width=1em, minimum height=2em,right=0cm of bf9] (bf10) {0};
		\node[draw,minimum width=1em, minimum height=2em,right=0cm of bf10] (bf11) {0};
		\node[draw,minimum width=1em, minimum height=2em,right=0cm of bf11] (bf12) {0};
		\node[draw,minimum width=1em, minimum height=2em,right=0cm of bf12] (bf13) {0};
		\node[draw,minimum width=1em, minimum height=2em,right=0cm of bf13] (bf14) {0};
		\node[draw,minimum width=1em, minimum height=2em,right=0cm of bf14] (bf15) {1};
		\node[draw,minimum width=1em, minimum height=2em,right=0cm of bf15] (bf16) {0};
		\node[draw,minimum width=1em, minimum height=2em,right=0cm of bf16] (bf17) {0};
		\node[draw,minimum width=1em, minimum height=2em,right=0cm of bf17] (bf18) {0};
	\end{tikzpicture}
	\begin{tikzpicture}
		\node[] (label0)                    {$h1'(s)=MD5(Random(10) + s)$};
		\node[below=.5cm of label0.west, anchor=west] (label1) {$h2'(s)=MD5(Random(10) + s)$};
		\node[below=.5cm of label1.west, anchor=west] (label2) {$Insert(\beta',password1234) \to \{9,14\}$};

		\node[draw,minimum width=1em, minimum height=2em,below=.7cm of label2.west, anchor=west] (bf1) {0};
		\node[draw,minimum width=1em, minimum height=2em,right=0cm of bf1] (bf2) {0};
		\node[draw,minimum width=1em, minimum height=2em,right=0cm of bf2] (bf3) {0};
		\node[draw,minimum width=1em, minimum height=2em,right=0cm of bf3] (bf4) {0};
		\node[draw,minimum width=1em, minimum height=2em,right=0cm of bf4] (bf5) {0};
		\node[draw,minimum width=1em, minimum height=2em,right=0cm of bf5] (bf6) {0};
		\node[draw,minimum width=1em, minimum height=2em,right=0cm of bf6] (bf7) {0};
		\node[draw,minimum width=1em, minimum height=2em,right=0cm of bf7] (bf8) {0};
		\node[draw,minimum width=1em, minimum height=2em,right=0cm of bf8] (bf9) {1};
		\node[draw,minimum width=1em, minimum height=2em,right=0cm of bf9] (bf10) {0};
		\node[draw,minimum width=1em, minimum height=2em,right=0cm of bf10] (bf11) {0};
		\node[draw,minimum width=1em, minimum height=2em,right=0cm of bf11] (bf12) {0};
		\node[draw,minimum width=1em, minimum height=2em,right=0cm of bf12] (bf13) {0};
		\node[draw,minimum width=1em, minimum height=2em,right=0cm of bf13] (bf14) {1};
		\node[draw,minimum width=1em, minimum height=2em,right=0cm of bf14] (bf15) {0};
		\node[draw,minimum width=1em, minimum height=2em,right=0cm of bf15] (bf16) {0};
		\node[draw,minimum width=1em, minimum height=2em,right=0cm of bf16] (bf17) {0};
		\node[draw,minimum width=1em, minimum height=2em,right=0cm of bf17] (bf18) {0};
	\end{tikzpicture}
	\begin{tikzpicture}
		\node[] (label0) {$h1''(s)=MD5(fixed salt_1 + s)$};
		\node[below=.5cm of label0.west, anchor=west] (label1) {$h2''(s)=MD5(fixed salt_2 + s)$};
		\node[below=.5cm of label1.west, anchor=west] (label2) {$Insert(\beta'',password1234) \to \{3,11\}$};

		\node[draw,minimum width=1em, minimum height=2em,below=.7cm of label2.west, anchor=west] (bf1) {0};
		\node[draw,minimum width=1em, minimum height=2em,right=0cm of bf1] (bf2) {0};
		\node[draw,minimum width=1em, minimum height=2em,right=0cm of bf2] (bf3) {1};
		\node[draw,minimum width=1em, minimum height=2em,right=0cm of bf3] (bf4) {0};
		\node[draw,minimum width=1em, minimum height=2em,right=0cm of bf4] (bf5) {0};
		\node[draw,minimum width=1em, minimum height=2em,right=0cm of bf5] (bf6) {0};
		\node[draw,minimum width=1em, minimum height=2em,right=0cm of bf6] (bf7) {0};
		\node[draw,minimum width=1em, minimum height=2em,right=0cm of bf7] (bf8) {0};
		\node[draw,minimum width=1em, minimum height=2em,right=0cm of bf8] (bf9) {0};
		\node[draw,minimum width=1em, minimum height=2em,right=0cm of bf9] (bf10) {0};
		\node[draw,minimum width=1em, minimum height=2em,right=0cm of bf10] (bf11) {1};
		\node[draw,minimum width=1em, minimum height=2em,right=0cm of bf11] (bf12) {0};
		\node[draw,minimum width=1em, minimum height=2em,right=0cm of bf12] (bf13) {0};
		\node[draw,minimum width=1em, minimum height=2em,right=0cm of bf13] (bf14) {0};
		\node[draw,minimum width=1em, minimum height=2em,right=0cm of bf14] (bf15) {0};
		\node[draw,minimum width=1em, minimum height=2em,right=0cm of bf15] (bf16) {0};
		\node[draw,minimum width=1em, minimum height=2em,right=0cm of bf16] (bf17) {0};
		\node[draw,minimum width=1em, minimum height=2em,right=0cm of bf17] (bf18) {0};
	\end{tikzpicture}
	\begin{tikzpicture}
		\node[] (label0) {$\widehat{h1}(s)=MD5(AES(k,fixed salt_1), s)$};
		\node[below=.5cm of label0.west, anchor=west] (label1)
		{$\widehat{h2}(s)=MD5(AES(k,fixed salt_2), s)$};
		\node[below=.5cm of label1.west, anchor=west] (label2)
		{$Insert(\widehat{\beta},password1234) \to \{5,9\}$};

		\node[draw,minimum width=1em, minimum height=2em,below=.7cm of label2.west, anchor=west] (bf1) {0};
		\node[draw,minimum width=1em, minimum height=2em,right=0cm of bf1] (bf2) {0};
		\node[draw,minimum width=1em, minimum height=2em,right=0cm of bf2] (bf3) {0};
		\node[draw,minimum width=1em, minimum height=2em,right=0cm of bf3] (bf4) {0};
		\node[draw,minimum width=1em, minimum height=2em,right=0cm of bf4] (bf5) {1};
		\node[draw,minimum width=1em, minimum height=2em,right=0cm of bf5] (bf6) {0};
		\node[draw,minimum width=1em, minimum height=2em,right=0cm of bf6] (bf7) {0};
		\node[draw,minimum width=1em, minimum height=2em,right=0cm of bf7] (bf8) {0};
		\node[draw,minimum width=1em, minimum height=2em,right=0cm of bf8] (bf9) {1};
		\node[draw,minimum width=1em, minimum height=2em,right=0cm of bf9] (bf10) {0};
		\node[draw,minimum width=1em, minimum height=2em,right=0cm of bf10] (bf11) {0};
		\node[draw,minimum width=1em, minimum height=2em,right=0cm of bf11] (bf12) {0};
		\node[draw,minimum width=1em, minimum height=2em,right=0cm of bf12] (bf13) {0};
		\node[draw,minimum width=1em, minimum height=2em,right=0cm of bf13] (bf14) {0};
		\node[draw,minimum width=1em, minimum height=2em,right=0cm of bf14] (bf15) {0};
		\node[draw,minimum width=1em, minimum height=2em,right=0cm of bf15] (bf16) {0};
		\node[draw,minimum width=1em, minimum height=2em,right=0cm of bf16] (bf17) {0};
		\node[draw,minimum width=1em, minimum height=2em,right=0cm of bf17] (bf18) {0};
	\end{tikzpicture}
	\caption{Different hash functions generation.  These
	functions will lead to different use-cases.  In the first two
	figures we have a fully random generation which leads to
	different cases every time as we cannot predict the output got
	from $Random$ function.  The fourth case employes a cryptographic
	function and a set of fixed strings ($fixed salt_n$) to generate
	the same set of hash function based on a secret key $k$.}
	\label{fig:hashing}
\end{figure}
\subsection{Anagram attack}
\label{sec:anagramatk}
The system as described in the previous sections is vulnerable to an
attack which exploits the order of the discovered n-grams.  This attack
aims to reconstruct the password as the anagram of the various n-grams.
The attack is composed of four steps:
\begin{enumerate}
	\item Generate all the hashes for a specific n-gram;
	\item Hash the n-grams into a Bloom filter;
	\item Analyze the Bloom filter and get the position of bits set to the true value;
	\item Compose the various n-grams to create a password.
\end{enumerate}
This scenario can be disruptive and can lead to the full
disclosure of hashed password in no time.  Also, this kind of attack can
be enhanced with the
help of a word dictionary similar to the one used in classical password attacks:
the search tree can be pruned by excluding the words which do not contain the
n-gram.
For instance, let us imagine that a user inserts the password ``password!!'' in the
filter with $\nu = 2$. Accordingly, the attacker will generate
all the possible bi-grams.  This, using an alphabet $\Delta$ will result in a generation
of $|\Delta|^2$ bi-grams which, for the ASCII case, is $(127-32)^2 = 9025$ bi-grams, an operation
which requires at most a couple of milliseconds on any modern CPU.  After this step,
the attacker will hash the bi-grams inserting them into the filter, which requires $\Theta(n)$
insertions with $n$ as the number of bi-grams.  Subsequently, the
attacker can create all the possible combinations in the search space generated by the pruned alphabet of bi-grams $\Delta_{II}$.  The research can be conducted by using an incremental number of repetitions. Therefore, the number of combinations which can be generated using the corresponding Bloom filter are:
\begin{equation}
{\Delta_\nu\choose \dfrac{n}{\nu}}
\end{equation}
with $n$ as the length of the searched string and $\nu$ the grade of the n-grams.  In the case of a common password of $8$ ASCII characters and a Bloom filter constructed with bi-grams the formula will result to:
\begin{equation}
{9025\choose \dfrac{8}{2}} \approx 2.76\times10^{14}
\end{equation}
combinations.  If we consider the worst case with repetitions the formula become
\begin{equation}
{\Delta_\nu + \dfrac{n}{\nu} - 1 \choose \dfrac{n}{\nu}}
\end{equation}
which, in this example, will result to ${ 9025 + 4 - 1 \choose 4}$ which is almost the same as the non-repetition case.  Passwords can be longer than $8$ char to provide enough security against brute-force or dictionary attacks.  For a password varying from $n$ characters to $N$ the number of combinations are:
\begin{equation}
\sum^{N}_{i=n} {\Delta_\nu + \dfrac{i}{\nu} - 1 \choose \dfrac{i}{\nu}}
\end{equation}
In our data-set the average password size was 11.56, therefore, limits between
8 and 14 can be evaluated resulting in:
\begin{equation}
\sum^{14}_{i=8} {9025 + \dfrac{i}{2} - 1 \choose \dfrac{i}{2}} \approx 9.96\times10^{23}
\end{equation}
The crypto-analysis of the attack should include the details of the filter like the
size of the bucket or the number of hash functions used.
\section{Experimental analysis}
\label{sec:experimental}
The specified method to check password similarity has been implemented
in C language.  The hash functions used was the standard
OpenSSL\footnote{https://www.openssl.org/} implementations of hash functions, in this case MD5.
The system was checked on a Ubuntu 18.04 system running in a VirtualBox virtual machine with
2 virtual CPUs and 1 GB of memory.  The hypervisor runs over an Intel core
i7-8700 cpu which clock frequency runs at 3.2GHz, and the host is used exclusively to run the test VM.
The random data was provided by
{\texttt~/dev/urandom} to avoid blocking behaviour\cite{gutterman2006analysis} and
read to generate random hash functions.  We used the following queries used to check the
filter:
\begin{enumerate}
	\item $\beta \leftarrow Create(\Gamma,\kappa)$
	\item $Insert(\beta,AAAA)$
	\item $Insert(\beta,BBBB)$
	\item $Check(\beta,AAAA)$
	\item $Check(\beta,CCCC)$
	\item $Check(\beta,BBBB)$
\end{enumerate}
\begin{figure}
	\includegraphics[scale=.7]{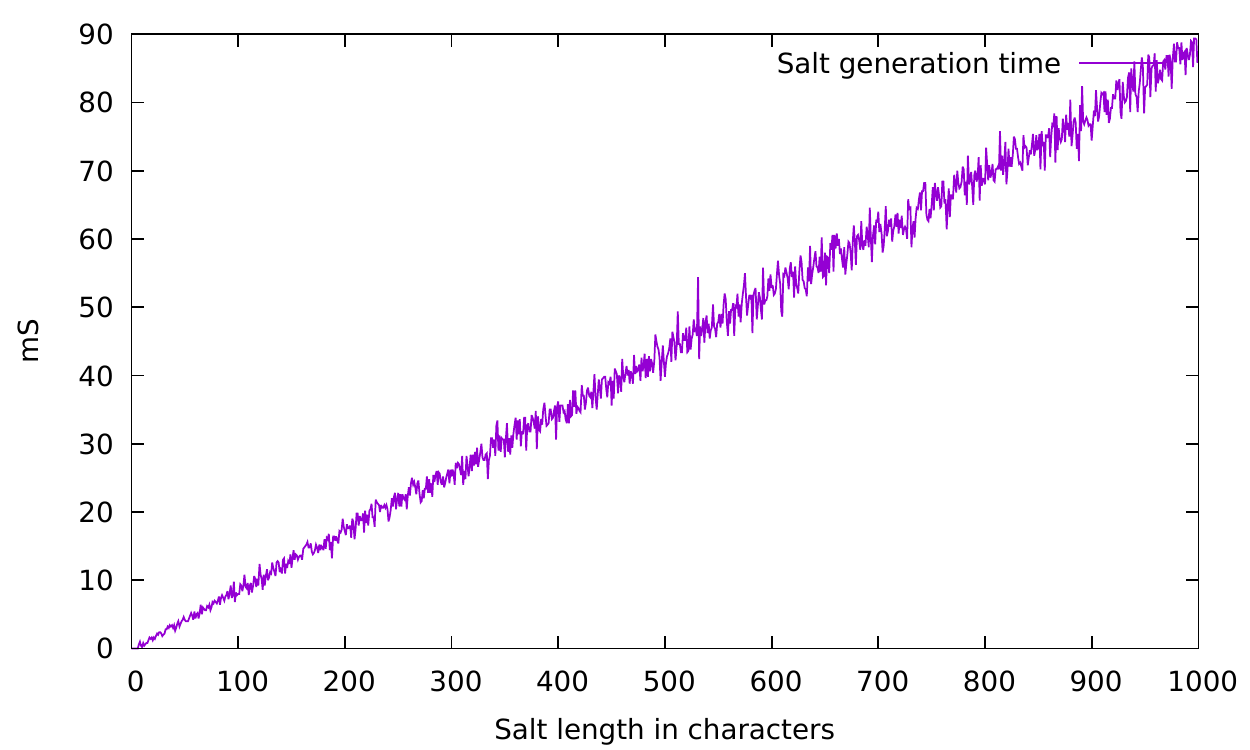}
	\caption{Creation time of the filter changing the size of the salt strings.}
	\label{fig:ex_salt_change}
\end{figure}
In this case the hash dimension seems not to influence the performances of
the filter. The computational load is dominated by the generation of the salt string.  The
performance of the salt string generation presented in figure \ref{fig:ex_salt_change} are
the averaged results of 5 runs of experiments, in which the size of the salt
changes from 1 (really easy to brute-force) to 1000 (really hard to
brute-force).  As shown in the graph, the performance
decreases linearly when the salt size increases.  This happens because a single
random character of the salt must be multiplied by the number of hash
functions present in the filter.
The $QInsert$ and $Distance$ performances has been evaluated using the
following querying pattern:
\begin{enumerate}
	\item $\beta_1 \leftarrow Create(\Gamma,\kappa)$
	\item $QInsert(\beta_1,thisismypassword,2)$
	\item $\beta_2 \leftarrow Create(\Gamma,\kappa)$
	\item $QInsert(\beta_2,thisismyp4ssword,2)$
	\item $\beta_3 \leftarrow Create(\Gamma,\kappa)$
	\item $QInsert(\beta_3,thisismypassw0rd,2)$
	\item $Distance(\beta_1,\beta_2)$
	\item $Distance(\beta_1,\beta_3)$
\end{enumerate}
In this test run, as in the Insert evaluation, the computational load
is dominated by the filter generation.
%
%
This system was also tested using c.a. 5000 credentials from real-world 
leaked institutional logins. Data were clustered and analyzed using the
filter, to observe which values are correctly identified as similar.
\begin{figure}
	\includegraphics[scale=.7]{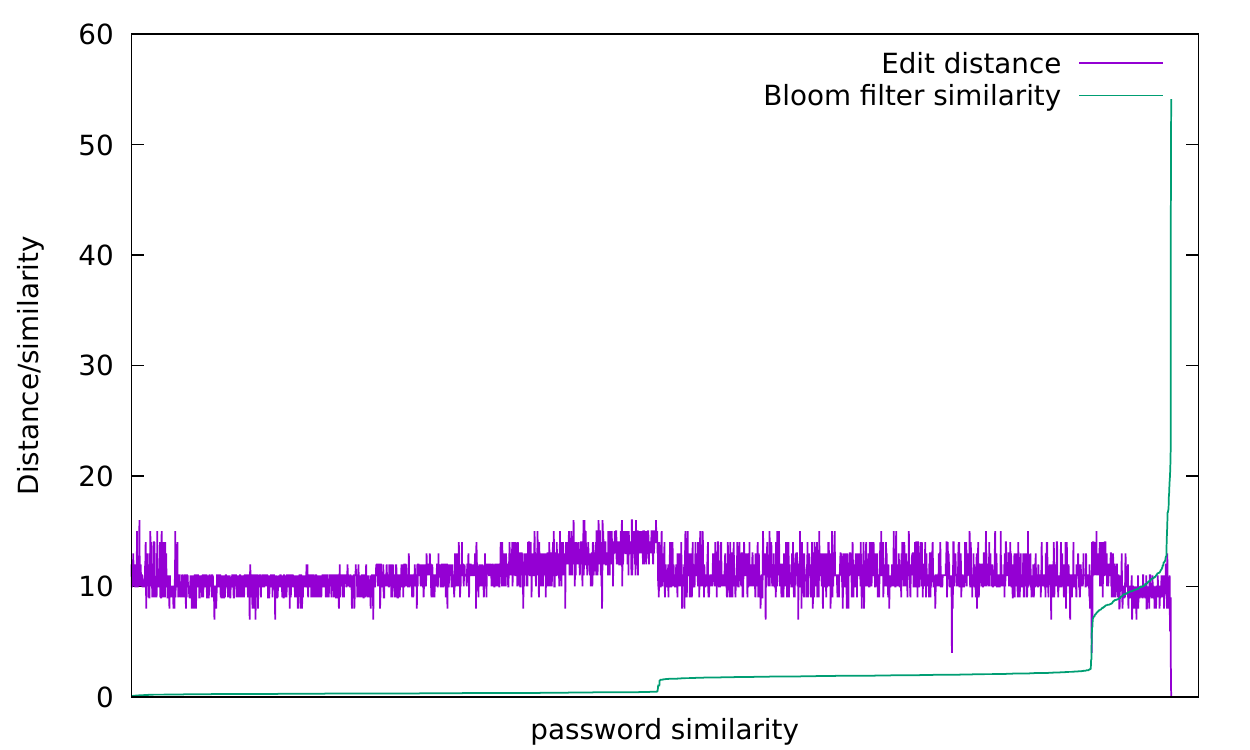}
	\caption{Performances of the filter compared to the edit distance applied to the password data set}
	\label{fig:ex_evaluation}
\end{figure}
The graph in figure \ref{fig:ex_evaluation} represent the analysis of
the filter's precision.   The graph also
shows the edit distance between the various strings.
\section{Application scenarios}
\label{sec:application}
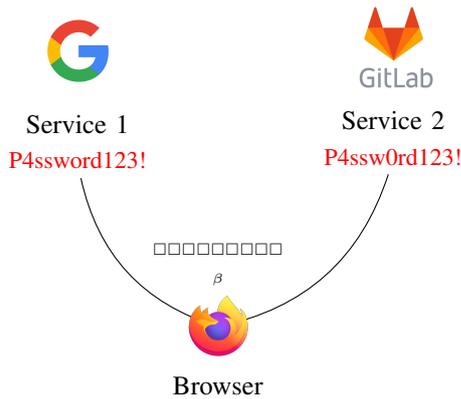
\begin{figure}
	\centering
	\begin{tikzpicture}
		\node[] (placeholder) {};
		\node[left=1cm of placeholder] (google)	{\includegraphics[scale=.15]{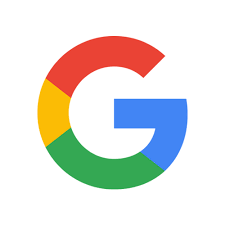}};
		\node[below=0cm of google] (label1) {Service 1};
		\node[below=0cm of label1] (password1) {\small\color{red}P4ssword123!};

		\node[right=1cm of placeholder] (gitlab)
		{\includegraphics[scale=.1]{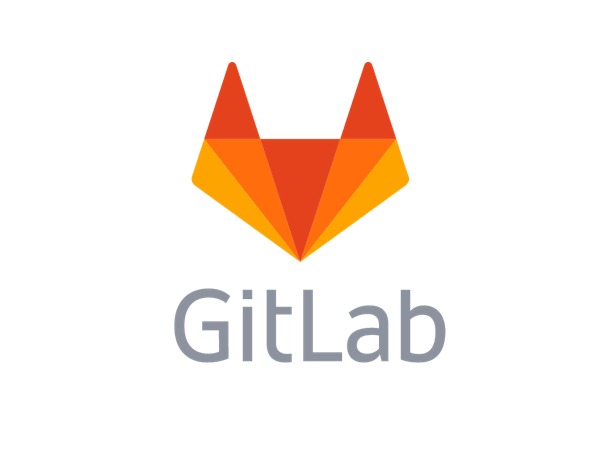}};
		\node[below=-.25cm of gitlab] (label2) {Service 2};
		\node[below=0cm of label2] (password2) {\small\color{red}P4ssw0rd123!};

		\node[below=3cm of placeholder] (browser)
		{\includegraphics[scale=.03]{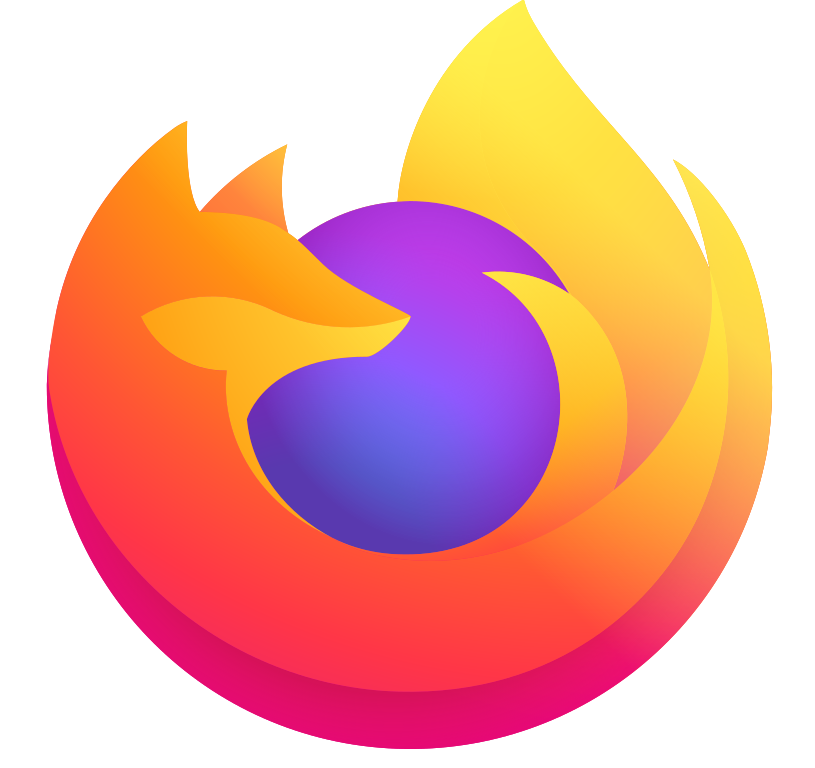}};
		\node[below=0cm of browser] {Browser};

		\node[above=.3cm of browser] (bf)
		{\tiny$\square\square\square\square\square\square\square\square\square$};
		\node[below=0cm of bf] (bf) {\tiny$\beta$};

		\begin{pgfonlayer}{bg}    
			\draw (browser.center) to[bend left]  (password1);
			\draw (browser.center) to[bend right] (password2);
		\end{pgfonlayer}
	\end{tikzpicture}
	\caption{Application scenario: The browser can insert the
	passwords $P4ssword123!$ and $P4ssw0rd123!$ into the Bloom
	filter, checking if they are similar enough to trow a warning of
	password similarity.}
	\label{fig:applicationfirefox}
\end{figure}

In this work, we described a technique to analyze password similarity maintaining
a good trade-off between utility and privacy. The proposed use case presents
a system that can notify if a new password is not different enough from a previous
one.
%
%
%
The typical application could be a browser's plug-in that issues a warning when the distance between a new password chosen by the user and the strings saved in the plug-in is below a pre-declared threshold.
The application can be exploited both to
discourage the use of similar passwords over time and to prevent their use over
different domains as pictured in figure
\ref{fig:applicationfirefox}.
The application can be instrumented to report the results of equations in section \ref{sec:password} 
regarding not only the similarity check result, but also the various parameters characterizing
the filter, to evaluate the quality of the classification process.
%
\subsection{Future works}%
\label{sec:future}
We are actually working on several improvements of the proposed password similarity
algorithm:
\begin{itemize}
	\item As stated in the section \ref{sec:password}, the filter can be
	generated using encrypted salts in conjunction with a strong cryptographic hash
	function\footnote{For example SHA3.}.  This approach can be employed to ensure that
	data inserted in the structure are analyzable only by the owner of the secret
	key. The crypto-analysis of the resultant system should be explored to create secure
	Bloom filters.

	\item A comparison with deep-learning based techniques can be introduced. This
	comparison should therefore include an analysis of the resistance against data-set
	reverse engineering. We argue that a neural network-based system needs a bigger data-set
	than our Bloom filter-based one, and that the former approach can be difficult  to analyze using black-box classifiers.

	\item The analysis of the crypto-system can be improved with a more in-depth
	comparison with privacy preserving techniques, such as $\delta$-presence or
	differential privacy.  As stated in section \ref{sec:application}, these approaches
	can suffer from the same issues affecting the deep-learning based one, i.e.,  the user cannot provide
	a data-set big enough to make the analysis valuable.
	
	\item Analysis of homomorphic encryption could lead to devise an encryption
	scheme to compute distances between encrypted strings using algorithms present in literature\cite{cheon2015homomorphic}.

\end{itemize}
We claim that these analyses can lead to the creation of an useful password checker,
which, while respecting user experience guidelines and security
best-practices, can signal to them dangerous similarities between their passwords.
\section{Conclusion}
In this paper we proposed a system which can help the analysis of similar
passwords keeping them obfuscated or saved in a secure enclosure.
This proposal tries to advise against the behaviour of password reuse, acting from
an application point of view, using a controllable method which is otherwise
difficult to achieve.
The system we proposed uses Bloom filters, a family of probabilistic data structures,
as the core elements of its design. This choice is driven by two properties
of these artifacts: the predictability of their behavior, and the deterministic
cryptanalysis that can be executed over them.
In conclusion, we claim that the
proposed method can be integrated as a modular component in any kind
of authentication system, to discourage the use of vulnerable passwords.

\bibliographystyle{plain}
\bibliography{password_similarity_using_probabilistic_data_structures}

\end{document}